\documentclass[a4paper, 12pt]{article}
\usepackage{graphicx} 
\usepackage{authblk} 
\usepackage[parfill]{parskip} 
\usepackage{comment}
\usepackage{inconsolata}
\usepackage{xurl}

\title{\texttt{tskit\_arg\_visualizer}: interactive plotting of ancestral recombination graphs}
\author[1]{James Kitchens}
\author[2]{Yan Wong}
\affil[1]{Department of Evolution \& Ecology and Center for Population Biology, University of California - Davis}
\affil[2]{Big Data Institute, Li Ka Shing Centre for Health Information and Discovery, University of Oxford}

\date{July 2025}

\begin{document}

\maketitle

\section*{Abstract}

\textbf{Summary:} Ancestral recombination graphs (ARGs) are a complete representation of the genetic relationships between recombining lineages and are of central importance in population genetics. Recent breakthroughs in simulation and inference methods have led to a surge of interest in ARGs. However, understanding how best to take advantage of the graphical structure of ARGs remains an open question for researchers. Here, we introduce \texttt{tskit\_arg\_visualizer}, a Python package for programmatically drawing ARGs using the interactive D3.js visualization library. We highlight the usefulness of this visualization tool for both teaching ARG concepts and exploring ARGs inferred from empirical datasets.

\textbf{Availability and implementation:} The latest stable version of \\ \texttt{tskit\_arg\_visualizer} is available through the Python Package Index (\url{https://pypi.org/project/tskit-arg-visualizer}, currently v0.1.0). Documentation and the development version of the package are found on GitHub (\url{https://github.com/kitchensjn/tskit_arg_visualizer}).

\textbf{Supplementary materials:} Methods for creating the example ARGs seen in Figure 1 can be found in the supplementary materials.

\textbf{Contact:} jkitchens@ucdavis.edu

\section{Introduction}

At each position in the genome, genetic relationships between sampled genomes can be described by an evolutionary (or “gene") tree. During meiosis, recombination generates mosaic chromosomes by bringing together regions with distinct inheritance histories, such that many different local trees can exist along the chromosome. However, neighboring local trees are usually highly correlated, sharing the majority of their nodes and edges in common \cite{hein_reconstructing_1990, mcvean_approximating_2005}. An “ancestral recombination graph" (ARG), first coined by Griffiths \cite{griffiths_two-locus_1991}, describes the set of local trees for a genomic region woven together into a graphical structure based on their shared branches (for a more complete definition of ARGs, see \cite{hudson_properties_1983} and \cite{wong_general_2024}). ARGs capture the observable patterns of inheritance that underlie the genetic diversity of the samples \cite{ralph_efficiently_2020}, making them an extremely data-rich object for genetic studies \cite{nielsen_inference_2025}. Thanks to major advances in ARG inference methods, ARG-based analyses can now be applied to many different empirical systems \cite{rasmussen_genome-wide_2014, speidel_method_2019, kelleher_inferring_2019, ignatieva_kwarg_2021, zhang_biobank-scale_2023, deng_robust_2024}. Coupled with the development of associated statistical methods, ARGs are currently being used to investigate a broad range of population genetics questions, including recognizing recombinant lineages \cite{ignatieva_kwarg_2021, zhan_towards_2023}, identifying selected loci \cite{stern_approximate_2019, vaughn_fast_2024}, and reconstructing complex demographic histories \cite{fan_likelihood-based_2025}. With this growth in interest \cite{lewanski_era_2024}, there is a pressing need for tools that support ARG research and lower the barrier of entry into this field. 

Visualization is critically important for teasing apart and communicating the complex relationships encoded in an ARG. This is often done by plotting local trees along a genome, regularly focusing on individual trees at loci of interest (e.g. \cite{hubisz_inference_2020}), or emphasizing the correlation between trees using colors or additional lines (e.g. \cite{rasmussen_espalier_2023}). However, this tree-based approach necessarily obscures how these trees fit into the larger ARG. Alternatively, as an ARG is a form of directed acyclic graph, the graph structure itself can be visualized. For example, classical depictions generally show a network of nodes organized by time, linked by vertical and horizontal edges, with the root at the top and samples at the bottom \cite{griffiths_two-locus_1991, griffiths_ancestral_1996, wiuf_ancestry_1999}. Although useful with small, hand-built examples \cite{rasmussen_genome-wide_2014, lewanski_era_2024}, the resulting visualization can be confusingly tangled and hard to interpret for larger empirical ARGs which often reflect highly reticulated ancestries. Moreover, this visualization strategy can make it hard to focus on specific genomic loci of interest, as local trees are more difficult to discern amid the larger structure of the graph.

Here, we present \texttt{tskit\_arg\_visualizer}, a graph-based visualizer which addresses the problems of tangling and local tree representation using modern software techniques. Force-directed simulation is used to untangle the graph, and interactive highlighting is used to reveal both local trees and the extent of genome regions spanned by edges and their associated mutations. In doing so, the visualizer draws the full ancestral structure, while simultaneously allowing the user to explore genetic relationships within local genomic regions. \texttt{tskit\_arg\_visualizer} adds to a growing list of tools aimed at teasing apart these data-rich graphs.

\section{Results}

\subsection{Implementation}

\texttt{tskit\_arg\_visualizer} is a Python package that leverages the D3.js visualization library to draw interactive ARGs in a browser based environment. When run from the command-line, this package launches a browser window; alternatively, plotting is possible directly inside a Jupyter Notebook. Figures can also be drawn inside Quarto presentations, providing interactivity when teaching about ARGs in a class or workshop setting.

\subsection{Model}

As the name suggests, \texttt{tskit\_arg\_visualizer} is designed to easily integrate into the \texttt{tskit} ecosystem of packages, and provides functions to convert from a \texttt{tskit.TreeSequence} in a manner that scales to millions of nodes and edges (as was required for Figure 1C). This package represents an ARG as a collection of tables for nodes, edges, mutations, and breakpoints, altogether referred to as a \texttt{D3ARG}. This mirrors the \texttt{tskit} tree sequence tabular format but with a focus on efficiency of visualization rather than efficiency of storage. For example, in addition to graph specific details such as node times and connections, tables contain styling information used to customize the visualization. Another key difference is that an edge in a \texttt{D3ARG} can contain disjoint chromosomal intervals, whereas a \texttt{tskit} tree sequence stores a separate edge per interval. Additionally, tables can be manually built without \texttt{tskit} by using the \texttt{pandas} Python package \cite{team_pandas-devpandas_2025} or importing from a previously saved JSON file.

\subsection{Visualizations}

\texttt{tskit\_arg\_visualizer} offers two methods for viewing an ARG:

\begin{itemize}
    \item \texttt{draw()}: displays the full graph to the screen (Figures 1A and 1B)
    \item \texttt{draw\_node()}: useful for larger ARGs, first filters to a subgraph around a focal node before displaying (Figure 1C)
\end{itemize}

These can be combined with two edge-drawing stylizations. The conventional “orthogonal" format, shown in Figure 1B, connects parent and child nodes via vertical and horizontal lines. It is particularly suitable for “full ARGs" \cite{wong_general_2024} with marked recombination nodes and where common ancestor nodes have a maximum of 2 children. The more flexible “line" stylization, shown in Figure 1C, connects each parent to its child via a straight line, which better accommodates ARGs containing nodes with many direct connections.

In all cases, a customized force-directed simulation is used to untangle graph edges \cite{eades_heuristic_1984}. Force-directed simulations balance a combination of “forces” to position nodes in a visually interpretable way – often where nodes repel one another but are held together by edges in the graph that act as springs. A major advantage of this algorithm is that it can limit the number of crossed lines without the need for scenario specific rules or knowledge of the graph’s shape. As the y-axis position of a node corresponds with its age, we fix this value and only allow the node’s x-axis position to be updated by the simulation. Force-directed simulations are likely to settle into local optima rather than finding the true minimum number of crossed lines, so the visualizer also allows nodes to be dragged left and right along the x-axis by the user, helping to further untangle the graph if needed. 

The visualizer provides many interactive features that use the chromosome bar drawn under each graph. Hovering over a region of this bar highlights its corresponding local tree embedded in the graph (shown stylistically in Figure 1B as solid vs dotted edges). Conversely, hovering over an edge of the graph highlights the genomic regions for which that edge covers. Mutations can be displayed on edges and colored arbitrarily; hovering over a mutation highlights its position on the genome bar, and hovering over an edge similarly highlights the genomic position of all the mutations on that edge (see Figure 1C).

All figures can be saved into a variety of file formats, including PNG, SVG, and JSON. The latter can be passed back into the visualizer with \texttt{draw\_D3()} to reproduce figures exactly.

\section{Conclusion}

\texttt{tskit\_arg\_visualizer} offers methods for programmatically plotting ARGs and provides many interactive features to help researchers explore these graphs. Visualizing ARGs, either in their entirety or by zooming in to specific substructures, can help researchers identify elements related to critical biological processes or concerns in data quality control. The visualizer has been a very useful tool for communicating about ARG concepts to a broader audience within the online \texttt{tskit} tutorials and during multiple conference workshops.

\section*{Acknowledgments}

We thank Graham Coop, Jerome Kelleher, and the \texttt{tskit} community for their suggestions and feedback throughout the development of this project. This work was supported by the National Institutes of Health (NIH R35 GM136290) and the National Science Foundation (NSF DISES 2307175).

\textbf{Conflict of Interest:} None declared.

\begin{figure}
    \centering
    \includegraphics[width=\textwidth]{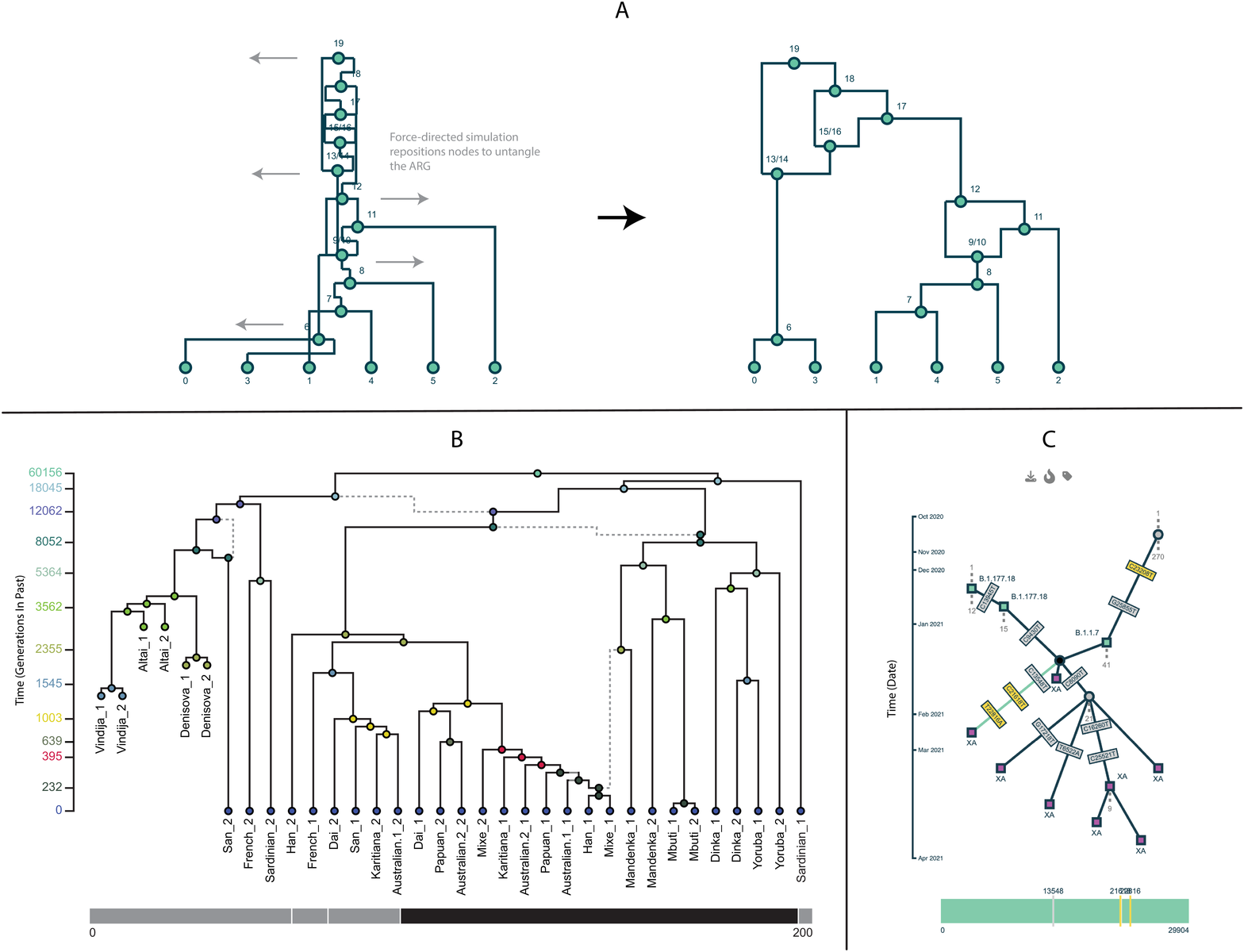}
    \caption{(A) The start and end layouts of a simulated ARG after nodes were repositioned by the force-directed simulation. (B) An inferred ancestral recombination graph covering 200bps of the human genome within the Duffy antigen receptor gene (DARC). This depicts the ancestry of 14 modern humans and 3 ancient samples. The local tree of SNP rs2814778 (1:159174683) is marked in black. Dashed lines indicate edges that do not appear in the local tree. The y-axis shows the number of generations in the past that events occurred. Nodes and edges positioned with \texttt{draw()}, exported as an SVG, and further stylized using Adobe Illustrator. (C) A SARS-CoV2 recombination subgraph for Pangolin XA recombinant genomes. This was produced by \texttt{draw\_node()} within a Jupyter Notebook. Viral samples are symbolized by squares, labeled with a Pangolin designation such as B.1.1.7. Dotted lines show connections to the remaining ~2.7 million ARG nodes. All 39 samples descending from the black recombination node are Pangolin XA recombinants; for brevity only 8 (in magenta) are shown, and all others descend via the dotted lines below the XA clade nodes. A nonlinear, ranked timescale has been used for efficient vertical node spacing. Mutations in the spike region have been plotted in gold; the edge leading to the left hand XA sample has been interactively highlighted, such that its mutations appear at their corresponding position on the lower genome bar. -- Further details regarding the construction and visualization of the ARGs in (B) and (C) can be found in the supplementary materials.}
\end{figure}

\section{Supplementary Materials}

\subsection{Creating Figure 1}

Code for each of the following subfigures can be found at \url{https://github.com/kitchensjn/tskit_arg_visualizer/docs/manuscript/Figure1}.

\subsubsection{A}

The ARG in this figure was simulated in with \texttt{msprime}. The version on the left was created by preventing the force-directed simulation from updating the node positions and saving the resulting ARG as an SVG. The versino on the right was saved after the force-directed simulation had settled into an optimum.

\subsubsection{B}

Following the ARGweaver tutorial from \cite{hubisz_inference_2020}, we sampled ARGs for 14 modern humans and 3 ancient samples around the DARC gene on chromosome 1. A random sampled ARG was trimmed to 100 basepairs on either side of the rs2814778 variant (1:159174583-159174783). This variant is found at high frequency in many African populations, excluding the San, and is believed to be under selection for malaria resistance \cite{mcmanus_population_2017}. The local tree for rs2814778 is very similar to that presented in Figure 5 from \cite{hubisz_inference_2020}, but it is now interwoven into the context of the surrounding trees. ARGweaver groups nodes into discrete time bins; for visual clarity, we separated the nodes by rank but colored them based on their true time (shown along the y-axis). The initial force-directed simulation untangled most of the graph, although some manually node positioning was required. The plot was downloaded as an SVG and brought into Adobe Illustrator to finish stylizing in preparation for a manuscript quality figure. For instance, though the visualizer provides a rudimentary y-axis by default, the y-axis in this figure was added post-hoc to better show the time discretization implemented by ARGweaver.

\subsubsection{C}

Figure 1C shows a small portion of a large ARG of 2,689,054 nodes, inferred using the sc2ts software \cite{zhan_towards_2023} from 2,482,157 coronavirus samples provided by the Viridian project \cite{hunt_addressing_2024}. Importing this ARG into the \texttt{tskit\_arg\_visualizer} took under a minute on a laptop. A subgraph of only 14 nodes is shown for readability: plotting larger subgraphs of hundreds of nodes is fully supported. The figure illustrates the earliest widely accepted SARS-CoV2 recombinant, labeled XA by the Pangolin project \cite{rambaut_dynamic_2020}, which is a combination of the B.1.177.18 lineage (to the left of the breakpoint) and the B.1.1.7 lineage (to the right). Mutations have been plotted using the default labeling scheme, showing the ancestral allelic state plus the genome position, plus the derived state. Here, we illustrate the bespoke coloring and labeling possibilities provided by the software, and the ability to highlight edges, which reveals the positions of the edge's mutations on the genome bar.

\bibliographystyle{plain}
\bibliography{main.bib}

\end{document}